\documentclass[conference]{IEEEtran}
\usepackage{pifont}
\usepackage{tabularx}
\usepackage{subfig}
\usepackage{booktabs}
\usepackage{pgfplots}
\usepackage{listings}
\usepackage{acronym}
\usepackage{csvsimple}
\usepackage{adjustbox}
\usepackage{xcolor}
\usepackage{float}
\usepackage{makecell}
\usepackage{tikz}
\usepackage{amsmath}
\usepackage{comment}
\usepackage{multirow}
\usepackage{cite}
\usepackage[colorinlistoftodos, textwidth=2.0cm]{todonotes}
\graphicspath{ {./images/} }
\usepackage{longtable}
\usepackage{xurl}
\usepackage{courier}
\usetikzlibrary{arrows.meta}
\usetikzlibrary{shapes}
\usetikzlibrary{external}
\usepackage{siunitx}
\DeclareSIUnit{\nothing}{\relax}
\DeclareSIUnit{\days}{days}
\usepackage{numprint}
\npdecimalsign{.}
\npthousandsep{,}

\usepackage{graphicx}
\newcommand{\Csharp}{%
  {\settoheight{\dimen0}{C}C\kern-.05em \resizebox{!}{\dimen0}{\raisebox{\depth}{\#}}}}
\newcommand{\Csharpspace}{%
  {\settoheight{\dimen0}{C}C\kern-.05em \resizebox{!}{\dimen0}{\raisebox{\depth}{\#}}} }

\usepackage{listings}
\usepackage{url}
\usepackage{breakurl}
\usepackage{xurl}
\usepackage[breaklinks]{hyperref}
\usepackage[flushleft]{threeparttable}
\usepackage{longtable}
\usepackage{amsfonts}
\usepackage{booktabs,tabularray}

\newcommand{\sys}{G-Scan }
\newcommand{\sysnospace}{G-Scan}

\usepackage{listings, xcolor}

\DeclareRobustCommand{\ttfamily}{\fontencoding{T1}\fontfamily{lmtt}\selectfont}

\definecolor{verylightgray}{rgb}{.97,.97,.97}

\lstdefinelanguage{Solidity}{
	keywords=[1]{anonymous, assembly, assert, balance, break, call, callcode, case, catch, class, constant, continue, constructor, contract, debugger, default, delegatecall, delete, do, else, emit, event, experimental, export, external, false, finally, for, function, gas, if, implements, import, in, indexed, instanceof, interface, internal, is, length, library, log0, log1, log2, log3, log4, memory, modifier, new, payable, pragma, private, protected, public, pure, push, require, return, returns, revert, selfdestruct, send, solidity, storage, struct, suicide, super, switch, then, this, throw, transfer, true, try, typeof, using, value, view, while, with, addmod, ecrecover, keccak256, mulmod, ripemd160, sha256, sha3}, 
	keywordstyle=[1]\color{blue}\bfseries,
	keywords=[2]{address, bool, byte, bytes, bytes1, bytes2, bytes3, bytes4, bytes5, bytes6, bytes7, bytes8, bytes9, bytes10, bytes11, bytes12, bytes13, bytes14, bytes15, bytes16, bytes17, bytes18, bytes19, bytes20, bytes21, bytes22, bytes23, bytes24, bytes25, bytes26, bytes27, bytes28, bytes29, bytes30, bytes31, bytes32, enum, int, int8, int16, int24, int32, int40, int48, int56, int64, int72, int80, int88, int96, int104, int112, int120, int128, int136, int144, int152, int160, int168, int176, int184, int192, int200, int208, int216, int224, int232, int240, int248, int256, mapping, string, uint, uint8, uint16, uint24, uint32, uint40, uint48, uint56, uint64, uint72, uint80, uint88, uint96, uint104, uint112, uint120, uint128, uint136, uint144, uint152, uint160, uint168, uint176, uint184, uint192, uint200, uint208, uint216, uint224, uint232, uint240, uint248, uint256, var, void, ether, finney, szabo, wei, days, hours, minutes, seconds, weeks, years},	
	keywordstyle=[2]\color{teal}\bfseries,
	keywords=[3]{block, blockhash, coinbase, difficulty, gaslimit, number, timestamp, msg, data, gas, sender, sig, value, now, tx, gasprice, origin},	
	keywordstyle=[3]\color{violet}\bfseries,
	identifierstyle=\color{black},
	sensitive=false,
	comment=[l]{//},
	morecomment=[s]{/*}{*/},
	commentstyle=\color{gray}\ttfamily,
	stringstyle=\color{red}\ttfamily,
	morestring=[b]',
	morestring=[b]"
}

\begin{document}
\title{\sysnospace: Graph Neural Networks for Line-Level Vulnerability Identification in Smart Contracts}

{\author
	{\IEEEauthorblockN{Christoph Sendner\IEEEauthorrefmark{1},
			Ruisi Zhang\IEEEauthorrefmark{2},
			Alexander Hefter\IEEEauthorrefmark{1},
			Alexandra Dmitrienko\IEEEauthorrefmark{1}, and
			Farinaz Koushanfar\IEEEauthorrefmark{2}}
		\IEEEauthorblockA{\IEEEauthorrefmark{1}University of W\"urzburg, Germany\\
		}
		\IEEEauthorblockA{\IEEEauthorrefmark{2}University of California San Diego, USA}
}}

\maketitle

\begin{abstract}
Due to the immutable and decentralized nature of Ethereum (ETH) platform, smart contracts are prone to security risks that can result in financial loss.
While existing machine learning-based vulnerability detection algorithms achieve high accuracy at the contract level, they require developers to manually inspect source code to locate bugs.
To this end, we present \sysnospace, the first end-to-end fine-grained line-level vulnerability detection system evaluated on the first-of-its-kind real world dataset. 
\sys first converts smart contracts to code graphs in a dependency and hierarchy preserving manner. Next, we train a graph neural network to identify vulnerable nodes and assess security risks. 
Finally, the code graphs with node vulnerability predictions are mapped back to the smart contracts for line-level localization. We train and evaluate \sys on a collected real world smart contracts dataset with line-level annotations on reentrancy vulnerability, one of the most common and severe types of smart contract vulnerabilities.
With the well-designed graph representation and high-quality dataset, \sys achieves 93.02\% F1-score in contract-level vulnerability detection and 93.69\% F1-score in line-level vulnerability localization. 
Additionally, the lightweight graph neural network enables \sys to localize vulnerabilities in 6.1k lines of code smart contract within 1.2 seconds.
\end{abstract}

\section{Introduction} 
\label{sec:introduction}

The recent success of Bitcoin~\cite{bitcoin} and Ethereum (ETH)~\cite{ethereumwhite} has brought decentralized platforms, which allow users to transact and interact in a trustless manner, to the forefront of attention.
Bitcoin, as the first generation blockchain platform, facilitates peer-to-peer digital currency transactions using proof-of-work consensus algorithms. However, due to the limited functionality of Bitcoin scripting languages, ETH and its smart contracts introduced a broader scenario where users can encode complex agreement terms directly into codes to enable automated and trustless transactions.

ETH has been widely applied in various domains, from financial services to supply chain management and beyond. Developers in the open-source ETH platform write their agreements into smart contracts code using Solidity and publish them on the blockchain to facilitate transactions with third parties. Like any other programming language, vulnerabilities also reside in ETH contracts, potentially leading to loss of funds or theft of sensitive information. The security risks arise from the two main aspects: First, due to the distributed nature of blockchain platforms, malicious third parties may publish their smart contracts to exploit vulnerabilities for profit. 
Second, as a newly developed programming language, Solidity has limited research on formal verification to ensure the smart contracts are written properly. Even minor mistakes in smart contract code can result in significant financial loss.

Many prior works explored vulnerability detection algorithms on Ethereum smart contracts, and most of them can be categorized into two types: (1) contract-level vulnerability detection and (2) line-level/node-level vulnerability detection. Contract-level vulnerability detection takes the detection as a classification problem and uses symbolic execution~\cite{Oyente,Mythril,Manticore}, fuzzing~\cite{ContractFuzzer,ConFuzzius,Echidna}, and machine learning algorithms~\cite{Peculiar,ContractWard,GNNpaper1,GNNpaper2,escort} to find vulnerable contracts. These methods achieve over 90\% F1 scores in classification, but developers are still required to examine the contract line-by-line to localize buggy statements. 

More recently, SCScan~\cite{SCScan} proposed a vulnerability detection system at the line level, while MANDO~\cite{Mando} proposed a node-level detection system with the aim of reducing the development burden. SCScan first uses a support vector machine (SVM) to detect vulnerable contracts and then a pattern matching algorithm to determine the exact positions of the vulnerabilities. However, SCScan is primitive which can only accept fix-size contracts as inputs to SVM, and the pattern matching algorithm fails to scale as new attacks appear. 
MANDO~\cite{Mando}, on the other hand, only detected vulnerabilities in the code graph but failed to map them back to the source code. MANDO employs a two-stage graph neural network (GNN) training to identify node-level vulnerabilities, where the first stage trains a graph classification model to assess the vulnerability of contracts, and the second stage utilizes node embeddings updated from a topology graph neural network to classify whether a specific node is vulnerable. MANDO can handle contracts of varying sizes, but after getting the node-level predictions, it cannot map vulnerabilities back to the smart contract for localization.  

Notwithstanding that node classification is widely studied by the machine learning community, localizing vulnerable nodes within the code graph itself is non-trivial and challenging for the following reasons. First of all, there is a lack of real-world datasets for fine-grained\footnote{throughout the paper, we use ``localize" and ``fine-grained detection" interchangeably.} vulnerability detection model training. MANDO~\cite{Mando} trains its GNN model on a simulated dataset by injecting vulnerabilities into clean contracts. These injections fail to represent some corner cases in real-world contracts and result in lower detection accuracy. 
Secondly, code graph construction in prior work~\cite{GNNpaper1,GNNpaper2,GNNpaper3} mainly extracts the semantic information in source code but fails to capture the code dependencies and hierarchies. 
Thirdly, as the contract grows, efficiently identifying vulnerable statements with regard to time and computation cost is crucial to save developers' time and resources.  

To tackle these challenges, we present \sysnospace, the first end-to-end fine-grained line-level vulnerability detection system with evaluations on the first-of-its-kind real world dataset. 
Our \sys converts the contracts to code graphs by first getting their abstract syntax tree (AST) representations and adding extra edges to reflect data dependencies and code hierarchies. Then, it assigns node features by including function types like \textit{While} and \textit{For}, variable properties like \textit{visibility} and \textit{storage location}, and node member access like \textit{send} and \textit{transfer}. Each node also processes a \textit{src} attribute which memorizes the lines where the nodes are extracted when constructing AST.  

\sys trains a GCN model on the converted graphs and learns the hidden features carrying the vulnerabilities. In the inference stage, \sys is used to predict the node labels of unseen code graphs and map back to the contract statements with \textit{src} attributes. 
We train and evaluate \sys on the first real-world fine-grained reentrancy vulnerability annotation dataset consisting of 13,773 smart contracts and achieve 93.02\% F1-score in contract-level vulnerability detection and 93.69\% F1-score in line-level vulnerability localization.

Our contributions to the community are summarized as follows:

\begin{itemize}
    \item We present \sysnospace, the first end-to-end algorithm that leverages graph neural networks for fine-grained line-level detection of smart contract vulnerabilities.
    \item \sys is a scalable and efficient framework that is agnostic to contract size and localizes vulnerability within 1.2 seconds for contracts with more than 6.1k lines of code.
    \item Proof of concept evaluation is demonstrated on the first-of-its-kind real world data collection consisting of 13,773 smart contracts and a total of 5,363,793 lines of code. Each data line is labeled to indicate whether they contain reentrancy vulnerabilities or not. 
     \item Our results show that \sys can achieve 93.02\% F1-score in contract-level vulnerability detection and 93.69\% F1-score in line-level vulnerability localization by incorporating variable dependencies and code hierarchies into graph modeling.

\end{itemize}

In summary, \sys provides an efficient and precise fine-grained vulnerability detection solution to smart contracts. \sys also contributes the first-of-its-kind large-scale real word smart contract dataset with fine-grained reentrancy vulnerability annotations. 
Our novel graph representation and lightweight GNN-based node classification model trained on the real-world dataset enable line-level vulnerability detection with both efficiency and accuracy compared with prior arts. We will open-source the code along with our collected dataset to promote research in this area. 

\noindent \textbf{Paper Organization:} For the rest of the paper, we will introduce vulnerability localization background and challenges in Section~\ref{sec:background}; the goal of \sys and the threat models in Section~\ref{sec:goal}; the detailed design pipeline of \sys in Section~\ref{sec:data}-Section~\ref{sec:training}; the experiments demonstrating \sysnospace's effectiveness in Section~\ref{sec:experiment}; and the conclusion and future work in Section~\ref{sec:conclusion}.
\section{Background and Challenges}
\label{sec:background}
\subsection{Smart Contracts Vulnerability}
Smart contracts are self-executing programs in Ethereum that enables developers to create decentralized blockchain applications, including financial services. These contracts are written in Solidity, a high-level programming language, and compiled into Ethereum Virtual Machine (EVM) bytecode for deployment on the Ethereum network.  Despite their benefits, smart contracts can also be vulnerable to security risks, which, if not identified in time, can lead to significant financial losses. In this context, we outline some of the vulnerabilities related to financial transactions, including the reentrancy attack, which is widely regarded as the most common and dangerous vulnerability (e.g., DAO~\cite{mehar2019understanding}).

\textbf{Reentrancy Attack:} The adversary performs reentrancy attacks~\cite{reentrancyvulnerability} by repeatedly calling back into a vulnerable contract before the previous invocation has been completed. It allows the adversary to repeatedly siphon funds.

\textbf{Integer Overflow/Underflow Attack:} The adversary performs overflow/underflow attacks~\cite{integerflow} on vulnerable smart contracts doing mathematical operations on unsigned integers. It allows the adversary to transfer excessive funds from the overflow/underflow.

\textbf{Access Control Attack:} The adversary performs access control attacks~\cite{accesscontrol} on vulnerable smart contracts that do not properly control access to sensitive functions or data. It allows the adversary to steal sensitive private information from the vulnerable smart contract.

\subsection{Graph Neural Networks}

Given a graph $G=(V,E,A)$ consists of a set of nodes $V = \{v_1, v_2,..., v_n\}$ and a set of edges $E = \{e_1, e_2,..., e_m\}$. For every node, $v_i \in V$ has a $k$ dimension feature vector $\mathbf{h}_i \in R^k$.  The connections between nodes are deposited into the adjacency matrix $A$ where a nonzero number in row $i$ and column $j$ means a connection from node $i$ to node $j$.

The objective of GNN is to learn a function $g$ that maps the feature embedding $\mathbf{h}_i$ of the $i$-th node $v_i$ to a new embedding vector $\hat{\mathbf{h_i}} \in R^h$ which capture the node's local and global information. For a multi-layer GNN model, the mapping above is performed iteratively to help nodes update their feature embeddings with the information from their neighbors via message passing. For every node $v_i$, it performs message passing by receiving the feature embeddings from its neighbors $j \in N_i$, where $N_i$ is the set of nodes adjacent to $v_i$. The messages are then aggregated using a customized function to obtain an updated representation $\hat{\mathbf{h_i}}$.

The computations are formulated in Equation~\ref{eq:GNN} where $f$, $g$, and $\oplus$ are customizable functions, e.g., convolution. Here, $\mathbf{h}_v^{(l)}$ and $\mathbf{h}_u^{(l)}$ are the node features at layer $l$, $\mathcal{N}(v)$ are the set of neighbors for node $v$, $\mathbf{e}_{u v}$ is the edge feature between nodes $u$ and $v$, and $\mathbf{h}_v^{(l+1)}$ is the updated features of node $v$ at layer $(l+1)$.
 \begin{equation}
 \label{eq:GNN}
 \mathbf{h}_v^{(l+1)}=g(\mathbf{h}_v^{(l)} \underset{u \in \mathcal{N}(v)}{\oplus} f(\mathbf{h}_u^{(l)}, \mathbf{h}_v^{(l)}, \mathbf{e}_{u v}))
\end{equation}

\subsection{Challenges}
To perform fine-grained line-level smart contract vulnerability detection, we identify the following challenges.

\textbf{Lack of Dataset: } Many real-world contract-level vulnerability detection datasets~\cite{Peculiar} are open-sourced for scientific research. However, these datasets lack line-level annotations that pinpoint the exact location of vulnerabilities. It brings challenges to the fine-grained detection model training, as high-quality data, like real-world contracts with annotations, is essential for the machine learning models to learn hidden patterns and achieve optimal performance.
 
\textbf{Graph Representation: } The second challenge is representing code graphs with both semantic and structural information. Previous approaches such as MANDO~\cite{Mando} and DR-GCN~\cite{GNNpaper2} extract heterogeneous code graphs based on semantic information from the source code, such as critical function calls/variables and temporal execution trace. However, they did not consider the structural information that indicates the code execution orders and relationships between data elements, which are where many vulnerabilities originate.
Apart from the edge connections, designing node feature embedding representations that depict function or expression types and properties helps GNN models learn better local and global code information. Therefore, developing appropriate graph representations is crucial in improving fine-grained vulnerability detection model performance.

\textbf{Efficiency: } Achieving efficiency in terms of inference time and cost is also challenging as the size of contracts scales. For time efficiency, line-level detection algorithms like MANDO~\cite{Mando} use a two-level detection algorithm, where it first classifies if a smart contract is vulnerable and then uses the node embeddings obtained from a topology graph neural network for line-level vulnerability classification. However, performing inference over three multi-level graph neural networks results in significant localization overheads. For cost efficiency, prior arts detect contract-level vulnerability via DNN based models~\cite{escort,GraphCodeBert,huang2018hunting} are parameter-heavy and require significant computation resources when contract users intend to verify if a specific contract is vulnerable. 
  
\textbf{Mapping between Node and Statement:} Another challenge is establishing the relationship between nodes in the code graph and statements in the smart contract. This is because localizing which line is vulnerable relies heavily on identifying the node representing the code block in the graph representation and accurately mapping it back to the corresponding line. While prior work such as MANDO~\cite{Mando} classifies vulnerable nodes within code graphs, they did not explicitly mention how the vulnerable nodes are mapped back line by line. In contrast, \sys proposed the first end-to-end vulnerability localization system by leveraging AST trees, which enables us to accurately map vulnerable nodes back to specific lines in the smart contract.
\section{Goal and Threat Model}
\label{sec:goal}

In this section, we will first introduce the goal of our proposed \sysnospace and then introduce the potential threat models that our \sys aims to defend. 

\subsection{Goal}
Our goal is to localize the vulnerabilities within smart contracts. While many works~\cite{Peculiar,ContractWard,GNNpaper1,GNNpaper2,escort} have used machine learning models to detect contract vulnerabilities, few explored how to localize them. 
Nevertheless, it is crucial to accurately and efficiently identify which line contains the vulnerability for both developers and end-users.
For developers, identifying the specific line of code containing the bug, instead of just contract-level detection and examining the code line-by-line, can improve working efficiency. Smart contracts can be inevitably complex, with intricate user agreements and transaction logics, making debugging a challenging task, even when developers know that vulnerabilities exist in the codebase. 
\sys helps the developers to focus on fixing the risky part instead of changing the entire codebase, which saves time and resources. 
For end users, with \sysnospace, fine-grained vulnerability detection with low inference overhead opens a more transparent door to facilitate them to analyze security concerns within contracts and avoid transactions with malicious parties.

In summary, \sys  contributes to improving the security of the transaction life-cycle from the following two aspects: 
\begin{itemize}
    \item Before publishing the smart contract, with \sysnospace, one can analyze the vulnerabilities within the smart contracts at line-level and fix them in time. 
    \item When making transactions with untrusted third parties, with \sysnospace, one can analyze the vulnerabilities within the smart contracts efficiently without costing too much computation resources.  
\end{itemize}

\subsection{Threat Model}

\begin{figure}[ht]
    \centering
    \includegraphics[width=0.8\columnwidth]{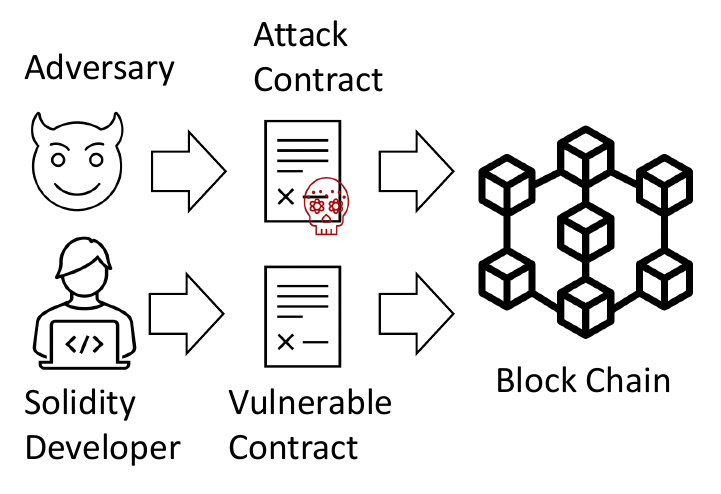}
    \caption{Attack Scenario. When solidity developer publishes a vulnerable contract to the blockchain, the adversary leverages another attack contract to exploit the vulnerabilities in the vulnerable contract and steal funds from the developer.}
    \label{fig:scenario}
\end{figure}

As depicted in Figure~\ref{fig:scenario}, a Solidity developer first publish a vulnerable contract to the blockchain. Then, an adversary exploits the vulnerabilities by publishing an attacking contract to call the vulnerable contract to steal funds. \sys safeguards smart contracts by helping Solidity developers identify and fix possible security risks promptly.

\textbf{Adversary’s Objective}
The adversary is the malicious party in the blockchain and attempts to perform an attack on the Ethereum platform. In this paper, we use the reentrancy attack as an example, but note that \sys can be adapted to detect any other vulnerability types.
In the reentrancy attack, the adversary deploys an attacking contract designed to exploit the vulnerabilities in the vulnerable contract. The vulnerable contract contains a function that allows it to transfer funds to other contracts. The adversary adds itself as one of the vulnerable contract recipients by first depositing the vulnerable contract funds and then withdrawing them. 
During the withdrawal execution, the attack contract repeatedly re-enters the vulnerable contract and calls the transfer function multiple times until all funds are stolen.

In Solidity, the above fund transfer is implemented by the following three subtype data structures, where $<$CA$>$ indicates the recipient's contract or account address and \texttt{x} describes the amount of fund to be transferred. 

\begin{itemize}
    \item \texttt{call}: \texttt{$<$CA$>$.call$\{$value: x$\}$}
    \item \texttt{transfer}: \texttt{$<$CA$>$.transfer(x)}
    \item \texttt{send}: \texttt{$<$CA$>$.send(x)}
\end{itemize}
 
The difference among the three subtypes mainly arises from their implementations: Firstly, the upper transaction limit of \texttt{transfer} or \texttt{send} is 2,300 gas (a unit to measure funds in ETH platform), while \texttt{call} has no such limitations. Therefore, if an adversary exploits the \texttt{call} subtype, it can result in a greater loss of funds. Secondly, \texttt{call} cannot automatically handle exceptions thrown by the called contract, which increases the potential for reentrancy attacks. Thirdly, due to the transaction limitations in \texttt{transfer} and \texttt{send}, some developers may opt to use \texttt{call} without mitigating the reentrancy vulnerabilities.

However, possible reentrancy attacks can still be performed on \texttt{transfer} and \texttt{send} for several reasons. Firstly, if these two subtypes requests call vulnerable contracts for funds exceeding the upper limit, exceptions can still happen, and if not handled properly, reentrancy attacks can be executed to steal funds. Secondly, if Ethereum Virtual Machine is upgraded and the maximum fund transaction limit changes, reentrancy attacks can still bring fund loss. Additionally, \texttt{transfer} will throw an exception if gas is depleted, leading to a state change reversal.

\textbf{Adversary’s Capacity}
We consider the most general attack case where the adversary is one of the contract creators. He or she has access to the public data structure on the chain and can upload his or her contract to the Ethereum system but does not have access to the detection of \sysnospace.

\begin{figure*}
    \centering
    \includegraphics[width=\linewidth]{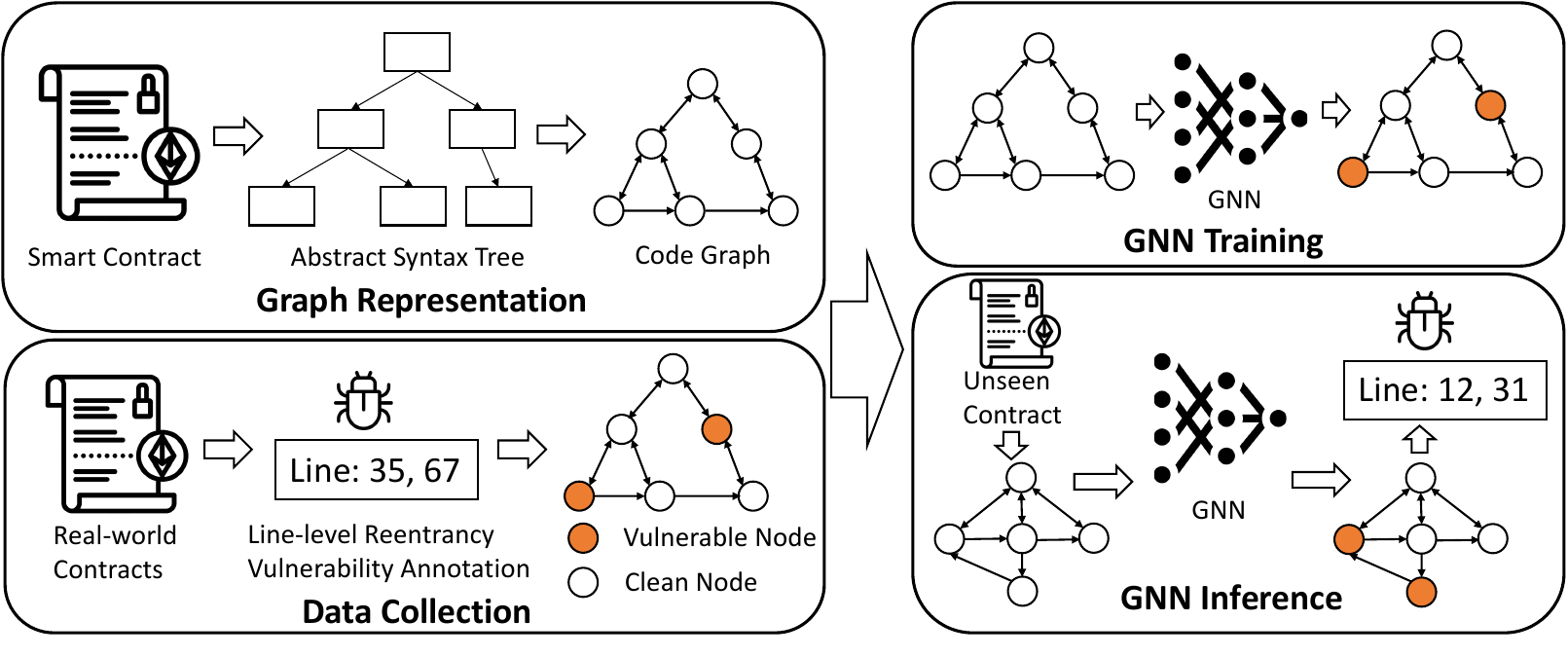}
    \caption{\sys overview.  First, we obtain the graph representation of the smart contract by converting it into an Abstract Syntax Tree (AST) representation and adding additional edges and node feature embeddings to reflect code hierarchies and dependencies. Next, we train a Graph Neural Network (GNN) model on the code graphs to classify vulnerable nodes and perform inference on unseen smart contracts. We train and evaluate the GNN model on a collected real-world smart contract dataset, with each data line annotated to indicate whether it contains reentrancy vulnerability.
    }
    \label{fig:global}
\end{figure*}

\section{Graph Representation}\label{sec:graph_construct}

In this and subsequent sections, we will first present how smart contracts are converted into code graphs, how to train GNN models to classify nodes as vulnerable or not, and then introduce how we collected our reentrancy vulnerability dataset, along with detailed statistics.

A more general pipeline of our \sys is shown in Figure~\ref{fig:global}. We begin by converting the smart contract into a code graph through AST representations that extract code structural information. We then add extra edges and node feature embeddings to reflect code hierarchies. Next, we train a GNN-based node classification model on the code graphs to classify vulnerable nodes and clean nodes. The training is performed on a collected real-world smart contract dataset with each data line annotated as either reentrancy vulnerable or not. 

The code graph representation construction consists of three steps, namely, AST generation, code graph edge generation, and node feature embedding generation. We will also illustrate how the code graph is constructed by giving an example in the end of this section.
   
\subsection{AST Generation} We use Solidity compiler~\cite{Soliditycompiler} to convert the smart contract into tree format, where the node represents syntactic constructs in the source code, such as statements and expressions, and each edge represents relationships between these constructs, such as function calls and value assignments.
The nodes within AST have two sets of node attributes: node type and source \texttt{src}. The node type describes the functionality of the source code, while the source \texttt{src} stores which part of the source code the node comes from. The edges in AST are not explicitly defined but are generated by the nested child nodes within each node.

\subsection{Code Graph Edge Generation} \label{subsec:add_edge}

In addition to the edges being implicitly defined by the hierarchically nested nodes in the AST, we also add the following edges to represent the structure and semantics of the underlying source code. These edges aim to obtain the representation by adding control flow, hierarchy between the nodes, and existing data dependencies between the nodes into the code graph. Specifically, we add the following edges into the code graph: (1) AST Hierarchy Edges; (2) Control Flow and Ordering Edges; (3) Reference Edges; (4) Branching Edges; (5) Loop Edges; and (6) Break, Continue, and Return Edges. After adding these edges to the code graph, we convert it into a directed homogenous graph for future classification in Section~\ref{sec:training}.

\begin{figure}[htbp]
  \centering
  \includegraphics[width=\linewidth]{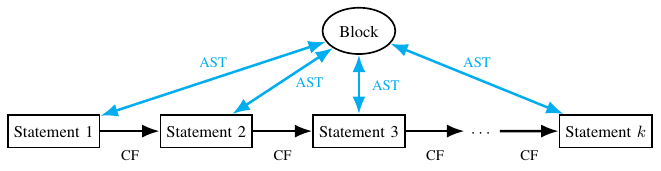}
    \caption{Control flow edges between k statements inside a block}
     \label{fig:control_flow}
\end{figure}

\textbf{AST Hierarchy Edges:} The first type of edges are the hierarchy edges generated by AST. Each node in the AST can have other child nodes defined in its node types, meaning the node and its child nodes have hierarchical relationships.
As shown in Figure~\ref{fig:control_flow}, for example, for node \texttt{Block}, it has node types related to \texttt{statements 1} to \texttt{statements k} and \texttt{Block} is the parent node of these child nodes. To represent the parent-child relationship in the code graph, we add two edges, one from the parent to the child and vice versa.  
  
\textbf{Control Flow and Ordering Edges:}  Apart from the hierarchy information defined by AST, we also add the control flow edges indicating the relationship between statements in the child nodes. There are three node types containing function statements, namely, \texttt{Block}, \texttt{UncheckedBlock}, and \texttt{YulBlock}. The function statements in the node types are given by the order they are executed in the original smart contracts. As shown in Figure~\ref{fig:control_flow}, for example, if a \texttt{Block} contains k statements, in addition to the AST edges added to indicate their hierarchy information, we also add edges between \texttt{statements} to explain the order they are executed. 

When a given node in the AST has the same child node belonging to two or more node types, ordering edges are introduced to distinguish each connection. For example, node type \texttt{Mapping} connects keys node $A$ to values node $B$, where $A$ and $B$ can be either the same or different. In its AST representation, the key node connects with the value node by attribute \texttt{keyType} defined in key node type.  The value node connects with its key node with \texttt{valueType} defined in the value node type.
When building AST, the connection is replaced by the undirected AST edges, which is hard to represent the mapping relationship between \texttt{keyType} and \texttt{valueType}. Therefore, we add an ordering edge to from \texttt{keyType} node to \texttt{valueType} node to represent the mapping relationship.

Apart from \texttt{Mapping}, we summarize other node types which also have similar ordering problems and how to add ordering edges in Table~\ref{tab:ordering}

\begin{table}[ht]
\centering
    \begin{tabular}{lcc}
    \toprule
 Node Type & Start & End\\
       \midrule
\texttt{IndexAccess}& \texttt{baseExpression} & \texttt{indexExpression} \\
  \multirow{2}{*}{\texttt{IndexRangeAccess}} & \texttt{baseExpression} & \texttt{startExpression} \\
    & \texttt{startExpression} & \texttt{endExpression}\\
    \texttt{FunctionCall} & \texttt{arguments} & \texttt{expression}\\
    \texttt{FunctionTypeName} &  \texttt{parameterTypes} &  \texttt{returnParameterTypes}\\
    \texttt{Assignment} & \texttt{leftHandSide} & \texttt{rightHandSide}\\
   \texttt{BinaryOperation} & \texttt{leftHandSide}& \texttt{rightHandSide}\\
    \texttt{FunctionDefinition} & \texttt{parameters} & \texttt{returnParameters} \\
   \texttt{YulFunctionDefinition} & \texttt{parameters} & \texttt{returnVariables}\\
    \bottomrule
    \end{tabular}
    \caption{Ordering edge directions\label{tab:ordering}}
  
\end{table}

\begin{figure*}[ht]
  \begin{center}
   \subfloat[If statement]{\includegraphics[width=0.19\linewidth]{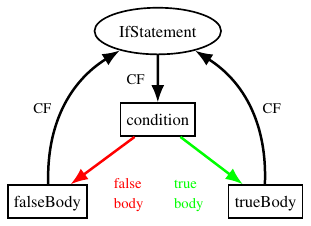}}
   \subfloat[Conditional]{\includegraphics[width=0.19\linewidth]{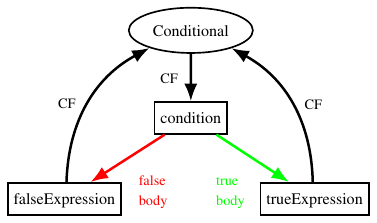}}
    \subfloat[If statement of Yul]{\includegraphics[width=0.19\linewidth]{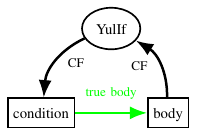}}
  \subfloat[Switch statement of Yul]{\includegraphics[width=0.19\linewidth]{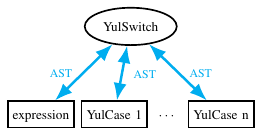}} 
  \subfloat[Case statement of Yul]{\includegraphics[width=0.19\linewidth]{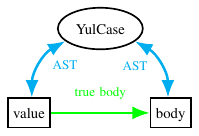}}
  \end{center}
  \caption{Graph structure of the node types referring to branchings}
  \label{fig:branch}
\vspace{-0.5cm}
\end{figure*}

\begin{figure*}[ht]
  \begin{center}
   \subfloat[For loop]{\includegraphics[width=0.24\linewidth]{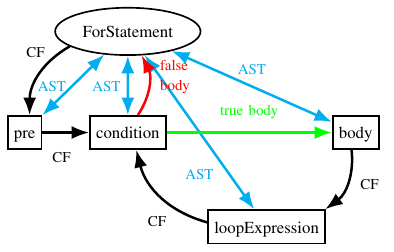}}
   \subfloat[Do while loop]{\includegraphics[width=0.24\linewidth]{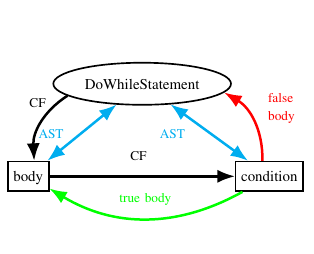}}
    \subfloat[While loop]{\includegraphics[width=0.24\linewidth]{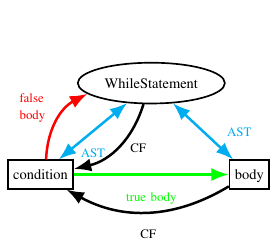}}
  \subfloat[For loop of Yul]{
  \label{fig:yul_loop}\includegraphics[width=0.24\linewidth]{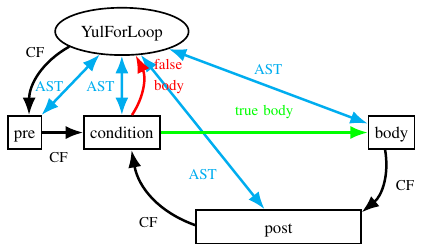}}    
  \end{center}
  \caption{Graph structure of the four loop types}
  \label{fig:loop_type}
  \vspace{-0.5cm}
\end{figure*}

\textbf{Reference Edges:} To ensure two smart contracts with the same functionality but different function or variable names have the same code graph, we introduce reference edges.
At variable level, AST introduced declaration node and assigned it a unique identifier attribute \texttt{id}. If the variables are again used in the smart contracts, the identifier nodes representing used variables will call the declaration node.
To reflect the data dependency between the usage of the same variable without a variable name, we introduce a new edge type called reference edge. The identifier node types are linked with their declaration node by the reference edge, which is always added in both directions, from the identifier node to the declaration node and vice versa.
Functions are treated in the same manner as variables, where reference edges are added from the identifier node to its declaration nodes and vice versa.

\textbf{Branching Edges:}
In the code graph representation, branching code blocks are only mapped to child nodes without further representation of the contexts between child nodes. To handle the branching, we add edges between the nodes representing the condition, the true branch, and the false branch.
Figure \ref{fig:branch} shows how branch conditions are handled for different functions. In the case of If statements and Conditional statements, three child nodes connect to the statement. One node represents the condition and connects the statement with control flow edges, one represents the true branch and connects the condition with \texttt{true body edges}, and the other represents the false branch and connects the condition with \texttt{false body edges}.

For If statements in Yul, which only have a true branch and ignore the statements in the true body for false conditions, we drop the \texttt{false body edges}. In the case of Switch statements in Yul, which consist of a switch statement with expressions and several case statements corresponding to the expressions, the switch statements are mapped by the AST edges between the node type  \texttt{YulSwitch} and case statement nodes of type \texttt{YulCase}.

For Case statements in Yul, AST edges are added between the \texttt{YulCase} node and the two child nodes at the node properties \texttt{body} and \texttt{value}. Additionally, a \texttt{true body edges} is added from \texttt{value} to \texttt{body} to reflect the body execution condition.

\textbf{Loop Edges:} We also include four loop edges that indicate variable dependencies in the code graph. The four loop types in Solidity are for, do-while, while, and for loop of Yul. Figure \ref{fig:loop_type} shows how the loop edges are added to the code graph.

For the For loop, there are four child nodes in the loop statement. When entering the loop, an initialization statement with loop parameters is executed and saved in the \texttt{pre} node. The initialization statement connects to the for statement via a control flow edge. Then, the loop condition is evaluated by the \texttt{condition} node by adding a control flow edge between the \texttt{pre} and \texttt{condition} nodes. If the evaluation result is false, the loop is exited, as reflected by a false body edge inserted from the \texttt{condition} node to the \texttt{ForStatement} node. If the evaluation result is true, the body node is executed, and a true body edge is added from the \texttt{condition} node to the \texttt{body} node. When the execution is finished, the loop is updated, and the loop condition is re-evaluated. To reflect this, we add two control flow edges, one from \texttt{body} node to the \texttt{loopExpression} node and one from the \texttt{loopExpression} node to the \texttt{condition} node.

In Do while loop and While loop, the \texttt{body} node is executed first, and then the \texttt{condition} is evaluated. Therefore, we add one false body edge from \texttt{condition} node either to the \texttt{DoWhileStatement} or to the \texttt{WhileStatement} node; and one true body edge from \texttt{condition} to the \texttt{body} node, and a control flow edge from the \texttt{body} to the \texttt{condition} node. We also add a control flow edge from either \texttt{DoWhileStatement} node and \texttt{WhileStatement} node  to the \texttt{condition} node meaning the entry point of the while loop. 

For the For loop of Yul, we add edges similar to the For loop. However, the attribute \texttt{post} gives the loop update here instead of the attribute \texttt{loopExpression}.

\textbf{Break, Continue, and Return Edges:} We also consider six node types that are related to leaving or continuing a loop:  \texttt{Break}, \texttt{Continue}, and \texttt{Return}, as well as their Yul counterparts \texttt{YulBreak}, \texttt{YulContinue}, and \texttt{YulLeave}. For \texttt{Break} and \texttt{YulBreak}, we add a control flow edge from the break statement node to the loop node, indicating the break enforces the node to leave the loop.  For \texttt{Continue} and \texttt{YulContinue}, we add a control flow edge from the continue statement node to the loop update node to show that the loop variable has been updated.
In the case of \texttt{YulLeave} or \texttt{Return}, we use control flow edges from the leave statement node to the function definition node to represent the loop has exited.

\subsection{Node Feature Embedding Generation}  \label{subsec:add_node}

The nodes in the code graph are derived from the AST, which provides a list of attributes that describe the main functionalities of each node and which part of the source code it comes from. Empirically, \cite{nodetypesAST} has identified 73 node types, but using one-hot encodings of all 73 types can introduce unnecessary computation overhead since some types have similar meanings and functions. Additionally, certain attributes may contribute more to the vulnerability than others.

To address these issues, we merged similar node types and created a 29-dimensional feature embedding. The first 20 dimensions capture information about different function or variable node definitions, the next eight dimensions describe node properties, and the final dimension indicates Solidity member access, specifically whether it involves \texttt{send} or \texttt{transfer}.

Table~\ref{tab:feat} displays some of the dimension information, where the first column denotes which dimension it comes from, the second column denotes the value of the specific node type it corresponds to in the third column, and the last column is additional descriptions. For the complete list of 29-dimensional vectors, please refer to the appendix.

\begin{table}[ht]
\centering
    \begin{tabular}{cclc}
    \toprule
 Dim & Values & Node Type & Description\\
       \midrule
 \multirow{4}{*}{0} & 1 & DoWhileStatement &  \multirow{4}{*}{Loop information} \\
 & 2 & WhileStatement & \\
 & 3 & ForStatement & \\
 & 4 & YulForLoop & \\
\midrule
\multirow{5}{*}{20} &  1 & 'internal' & \multirow{5}{*}{Different values of \texttt{visibility}}\\
& 2 & 'external' & \\
& 3 & 'private' & \\
& 4 & 'public' & \\
& 5 & unknown value & \\
\midrule
 \multirow{2}{*}{28} & 1 & 'transfer &  \multirow{2}{*}{Member access of attribute \texttt{memberName}}\\
 & -1 & 'send' & \\
    \bottomrule
    \end{tabular}
    \caption{Description of the node feature vectors\label{tab:feat}}
\end{table}

\textbf{Example}
We demonstrate the process of generating the code graph shown in Figure~\ref{fig:yul_loop} from the Solidity code below. The function's purpose is to withdraw a specified amount ( \texttt{\_amount}) of money from the ExampleYul contract and transfer it to a specified target address (\texttt{\_to}).

\begin{lstlisting}[language=Solidity, numbers=none]
pragma solidity ^0.8.19;
contract ExampleYul {
  function withdrawYul(address _to, uint256 _amount) external payable {
    assembly { 
    for {let i := 0} lt(i, 2) {i := add(i, 1)} {
        // loop body
    }
    }
    //other solidity text
  }
}
\end{lstlisting}

In the Yul Loop, variable $i$ is initialized to 0 using \texttt{let i := 0}; if the condition \texttt{lt(i, 2)} is satisfied, we increment $i$ by 1 using \texttt{i := add(i, 1)}. When $i$ satisfies the condition, the loop body is executed.

To generate the code graph from this smart contract, we extract the following nodes from the AST: \texttt{YulForLoop}, which defines the loop function; \texttt{pre}, which initializes the variable $i$; \texttt{condition}, which determines when to exit the loop in \texttt{lt(i, 2)}; and \texttt{post}, which updates $i$ using \texttt{i := add(i, 1)}. The loop body is represented by the \texttt{body} node.

Next, we add the following edges to the code graph. The first set of blue edges represents the parent-child relationships in the AST. Then, we add black control flow edges to indicate how the code is executed. First, the \texttt{pre} node initializes the variable $i$, followed by checking if $i$ satisfies the loop condition node \texttt{condition}. When the loop body represented by the \texttt{body} node is executed, we update $i$ following \texttt{post}, and then check if the loop condition node \texttt{condition} is still satisfied. The red and green loop edges connect the control flow edges. If the \texttt{condition} node is true, a true body edge goes from the \texttt{condition} node to the \texttt{body} node to continue the loop. If the \texttt{condition} node is false, a false body edge goes from the \texttt{condition} node to the \texttt{YulForLoop} node to exit the loop. Since there are no branch, return, variable, and function definitions in the code, we do not add these edges to the code graph.

\section{GNN Training and Inference}
\label{sec:training}
In this section, we present how we use GNN models to classify vulnerable nodes within code graphs. We train a GNN-based node classification model on the code graphs with annotated ground truth to predict vulnerabilities. The details of how we acquire the dataset are summarized in Section~\ref{sec:data}.

\subsection{Node Classification Training} 
The objective of the node classification model is to determine whether a given node is vulnerable.
We train a seven-layer GCN model to predict the node label, where $1$ means the node is vulnerable and $0$ means the node is clean.
The GNN model is trained by minimizing the cross-entropy between predicted labels and ground truth, as shown in Equation~\ref{eq:gnn_loss}. $N$ is the total number of classes, $G_i$ is the binary indicator (0 or 1) if class $i$ is the correct classification for this code graph, and $P_i$ is the predicted probability that class $i$ is the correct classification for this code graph.

\begin{equation}\label{eq:gnn_loss}
\mathcal{L}(G,P) = - \sum_{i=1}^{N} G_i \log(P_i)
\end{equation}

The detailed model architecture is summarized in Table~\ref{tab:GNNarchi}, which takes the contract code graph as input and predicts the binary labels for each node. The GNN model consists of seven GCN layers and a ReLU activation function after each layer. Then, the updated feature embeddings are passed through three linear layers, followed by a Softmax activation function to generate binary label masks. 

\begin{table}[htbp]
  \centering
  \begin{tabular}{cccc}
    \toprule
    Layers & Input Size & Output Size & Activation Function\\
    \midrule
    GCNConv0 & 29 & 500 & ReLU\\
    GCNConv1 & 500 & 500 & ReLU\\
    GCNConv2 & 500 & 500 & ReLU\\
    GCNConv3 & 500 & 500 & ReLU\\
    GCNConv4 & 500 & 500 & ReLU\\
    GCNConv5 & 500 & 500 & ReLU\\
    GCNConv6 & 500 & 500 & ReLU\\
    Linear0  & 500 & 300 & ReLU\\
    Linear1  & 300 & 100 & ReLU\\
    Linear2  & 100 & 2 & Softmax\\
    \bottomrule  
  \end{tabular}
  \caption{GNN architecture\label{tab:GNNarchi}}
\end{table}

\subsection{Node Classification Inference} 

In the inference stage, an unseen smart contract is first converted into an AST representation using the Solidity compiler. Then, we add additional edges and node features to form a code graph following the process outlined in Section~\ref{sec:graph_construct}. Next, the code graph is fed into the trained GNN model for inference. The final Linear2 layer in Table~\ref{tab:GNNarchi} predicts the labels for each node. Once we obtain the predicted node labels in the code graph, we map them back to the original smart contract using the node's \texttt{src} attributes, which enable us to determine which line the node corresponds to. If a node is labeled as vulnerable, the corresponding line is also marked as vulnerable.

\section{Data Collection}
\label{sec:data}

This section presents how we collect smart contracts and annotate the reentrancy vulnerabilities on each data line.  We also provide  visualizations of the dataset statistics.

\subsection{Dataset Construction}
Our ETH smart contracts are constructed using a previous version of the SmartBugs Wild Dataset~\cite{smartbugswild} called Peculiar~\cite{Peculiar}. The dataset construction process involved two stages. In the first stage, we clean the collected smart contracts and split them into training, validation, and test subsets. In the second stage, we annotate each data line in the smart contracts to identify the presence of reentrancy vulnerabilities.

\textbf{Preprocessing} Although the previous Peculiar dataset underwent cleaning, the duplication rate remains over 50\%. These can occur in two ways: (1) one contract may have been copied from another by adding a few white lines or comments; and (2) one contract may have been copied from another, with changes only made to variable names, function names, or variable values. Moreover, many contracts lacked their reference files, making it impossible to construct an AST tree.

To address these issues, we adopted a two-step approach. First, we generated an Abstract Syntax Tree (AST) tree from the Solidity compiler, which removed white space characters and assigned comment lines to an ignored node type. Next, we constructed the Solidity code graph by leveraging information on the smart contract code structure and the AST tree (refer to Section~\ref{sec:graph_construct} for details on graph construction). This step removed intermediate values, as well as variable and function names.
After obtaining the code graphs, we grouped the smart contract codes based on the similarity of their sha256 hashes. We selected one representative contract from each group, resulting in a total of 22,237 smart contracts, which are less than half of the original dataset containing 46,057 contracts. 

\textbf{Annotation} In the Peculiar dataset~\cite{Peculiar}, reentrancy vulnerability was only annotated at the contract-level for the \texttt{call} subtype. To achieve more accurate line-level annotation, we expanded the annotation from the \texttt{call} subtype to include all three subtypes: \texttt{call}, \texttt{send}, and \texttt{transfer}. Contracts that did not include these three subtype functions were deemed non-vulnerable.

In the remaining contracts, we manually inspected the data lines containing the three subtype functions to determine if a state change was made after the money transfer, and if there was no reentrancy lock in place. If these conditions were met, we annotated the lines as vulnerable. We then mapped these annotations to the code graphs. Each node in the code graph was assigned an attribute \texttt{src}, indicating the original smart contract line from which the node was derived. If the corresponding code line was marked as vulnerable, the node was labeled as such; otherwise, it was labeled as non-vulnerable. The final reentrancy vulnerability dataset only contains smart contracts with three subtype functions with 13,773 smart contracts.

\subsection{Dataset Statistics}

The dataset is split into training, validation, and test sets, with details provided in Table~\ref{tab:dataset}. In the table, the acronym VG represents the number of vulnerable graphs, which corresponds to the number of contracts in each dataset. Similarly, the acronym VN represents the total number of vulnerable nodes within the corresponding Solidity code graphs.

\begin{table}[ht]
\centering
    \begin{tabular}{cccccc}
    \toprule
 Subtype & Dataset & VG & non-VG  & VN & non-VN\\
       \midrule
 \multirow{3}{*}{Call} & training set & 573 & 11,040 & 28,900 & 11,468,676\\
   & validation set & 121 & 959 & 5,430 & 1,447,293\\
   & test set & 118 & 962 & 6,603 & 1,401,601\\
   \midrule
  \multirow{3}{*}{Send} & training set & 1,067 & 10,546 & 61,508 & 11,436,068\\
   & validation set & 200 & 880 & 10,678 & 1,442,045\\
   & test set & 226 & 854 & 13,207 & 1,394,997\\
   \midrule
  \multirow{3}{*}{Transfer} & training set & 1,572 & 10,041 & 83,469 & 11,414,107\\
  &  validation set & 320 & 760 & 16,782 & 1,435,941\\
  &  test set & 346 & 734 & 18,213 & 1,389,991\\
    \bottomrule
    \end{tabular}
    \caption{Dataset statistics on each subtype, VG represents vulnerable graphs number, VN represents vulnerable nodes number\label{tab:dataset}}
\end{table}
 
\begin{figure*}[ht]
  \begin{center}
   \subfloat[Node Number Distribution]{\includegraphics[width=0.3\linewidth]{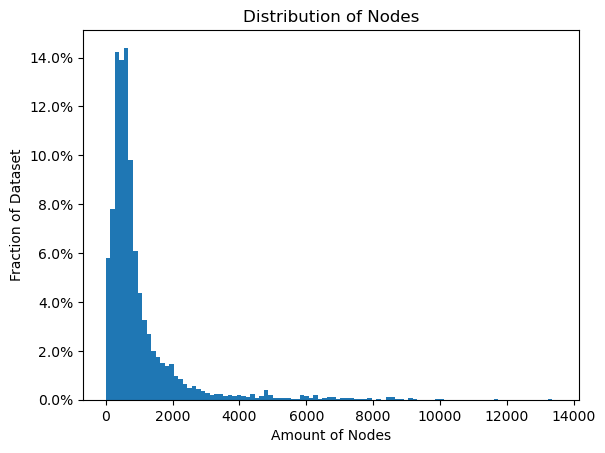}}
   \subfloat[Edges Number Distribution]{\includegraphics[width=0.3\linewidth]{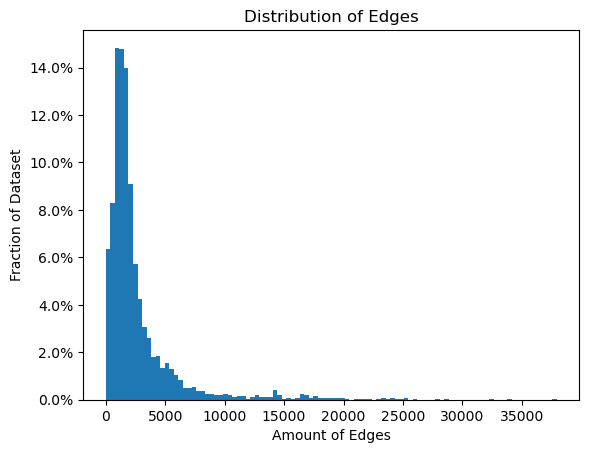}}
    \subfloat[Lines of Code Distribution]{\includegraphics[width=0.3\linewidth]{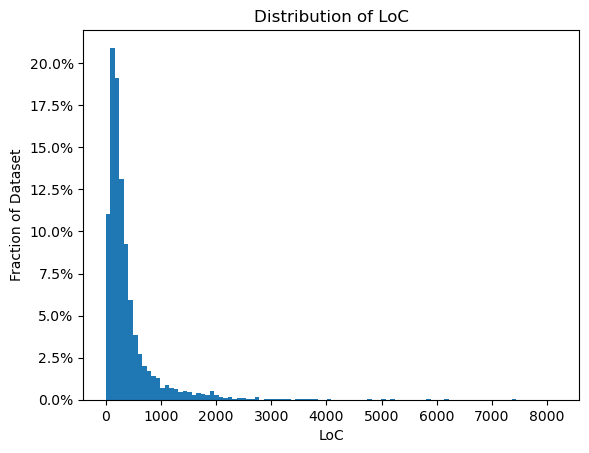}}
  \end{center}
  \caption{\sys dataset visualization on code graph node number, edge number, and smart contract lines of code}
  \label{fig:dataset_statistic}
  \vspace{-0.5cm}
\end{figure*}

In addition to the statistical information, we also present visualizations to demonstrate the code graph data distribution. Figure~\ref{fig:dataset_statistic} displays the code length distribution of smart contracts, as well as the nodes and edges distribution of each code graph. The first and second figure provides insights into the size and complexity of the code graphs, as measured by the number of nodes and edges, respectively. The third figure shows the distribution of contract length in terms of the number of lines of code.

\section{Implementation}
In this section, we present the general workflow for implementing \sys at both training and inference time.

\subsection{\sys Training Implementation}\label{sec:trainimple} 
The pipeline for our approach consists of three main steps: (1) constructing an AST from the source files; (2) converting the AST into a code graph with vulnerability labels; and (3) training a GNN-based node classification model. In the following subsection, we provide a detailed description of how each step is implemented.

\textbf{AST Construction} We construct ASTs from source code files using Solc~\cite{Soliditycompiler}. The AST construction requires the smart contracts in one file. Therefore, if the smart contract is written in multiple files, we merge them into a single file before performing AST conversion. The following is a sample Solidity program, where x.x.xx represents the Solidity version in which the contract was written.
It is worth noting that ASTs generated by Solc versions below 0.4.12 are not compatible with newer versions, and only a small number of contracts are written in these older versions. Therefore, we did not include them in our dataset collection.

\begin{lstlisting}[language=Solidity, numbers=none]
pragma solidity x.x.xx;
\end{lstlisting} 

We then convert the line-level annotations from smart contracts to AST node labels, as shown in the annotateLabel function below. It takes AST's node attributes \texttt{src} list 
 (\texttt{node2sourceCode}), smart contract's source code annotation list (\texttt{GTlabel}), and the source code at byte level (\texttt{codeBit}) as input. The \texttt{GTlabel} variable stores which line in the original smart contracts are vulnerable.

\lstset{basicstyle=\footnotesize}
\begin{lstlisting}[language=python, numbers=none]
def annotateLabel(node2sourceCode,GTlabel,codeBit):
    labels=[]
    if sum(GTlabel)==0:
        return [0]*len(node2sourceCode)

    for src in node2sourceCode:
        src=src.split(":")
        sBit=codeBit[:int(src[0])]
        eBit=codeBit[:(sBit+int(src[1]))]

        start=len(sBit.decode("utf-8").splitlines())
        end=len(eBit.decode("utf-8").splitlines())
        codeLine=[*range(start, end+1,1)]
        lineLabel=[GTlabel[i] for i in codeLine]

        if sum(lineLabel)>0:
            labels.append(1)
        else:
            labels.append(0)

    return labels
\end{lstlisting}

If there are no vulnerable annotations in the current smart contract, we return a set of 0 to indicate that the contract is clean. However, if the contract has vulnerable annotations, we proceed to analyze each node in the AST representation to obtain node-level labels.
Since the \texttt{src} attribute in the AST representation points from each node to the corresponding position in the original contract at the byte level instead of the line number, we need to decode the code byte into utf-8 format to determine the exact line number. We calculate the decoded line length and use this information to label each node as vulnerable (labeled as 1) if one of the corresponding lines is marked as vulnerable in the \texttt{GTlabel}. If none of the corresponding lines are marked as vulnerable, the node is labeled as not vulnerable (labeled as 0).

\textbf{Code Graph Generation} 
The ASTs are then converted to code graphs as described in Section~\ref{sec:graph_construct}. To generate the code graph, we use the following functions:

\begin{lstlisting}[language=Python, numbers=none]
from utils import dicToNodes 
from graph_construct import createReferenceEdges, createNodeFeatureVectors

graph = dicToNodes(AST)
graph = createReferenceEdges(graph, AST)
graph = createNodeFeatureVectors(graph, AST)
\end{lstlisting}
 
The \texttt{dicToNodes} function is used to extract nodes and edges from the AST and form the initial code graph. The AST object contains a dictionary with node mappings to the source code, as well as node and edge lists, edge types lists, and a list indicating breaks, continues, or returns in substructures. The graph object, on the other hand, includes a node mapping from the code graph id to the AST node id, as well as node and edge lists. The resulting code graph is stored in a PyG data object to facilitate future GNN training.

After creating the initial AST tree, we use the \texttt{createReferenceEdges} function to add additional edges to the code graph as described in Section~\ref{subsec:add_edge}.
We also add additional node features to the code graph using the \texttt{createNodeFeatureVectors} function, which takes the code graph and AST node properties as inputs and adds node features as described in Section~\ref{subsec:add_node}.

\textbf{GNN Training} The GNN model for node classification is trained with the code graphs in the training set and evaluated with the code graphs from the validation and test set. We add additional training details, including the hyperparameters in Table~\ref{tab:hyper}. The architecture information is summarized in Table~\ref{tab:GNNarchi}.

\begin{table}[htbp]
  \centering
  \begin{tabular}{lc}
    \toprule
    Variables & Settings\\
    \midrule
    Epoch & 1600 \\
    Batch size & 100 \\
    Optimizer & adam\\
    Learning rate & 0.0001\\
    Loss function & cross entropy\\
    \bottomrule  
  \end{tabular}
  \caption{GNN training hyperparameters\label{tab:hyper}}
\end{table}

\subsection{\sys Inference Implementation}
 
We utilize the trained GNN for node classification to predict the location of vulnerabilities. As with the training process for \sysnospace, we first convert the source code file into an AST representation using the Solidity compiler, and then transform it into a code graph representation without additional labels indicating vulnerability.
We then feed the code graph into the GNN model for node classification inference and obtain a binary mask that measures whether each node is vulnerable or not. If none of the nodes are vulnerable, we predict that the contract is clean. Otherwise, we predict which nodes are vulnerable.
After obtaining the node prediction results, we convert them back to code lines using the \texttt{src} node property and mark the vulnerable lines on the source code file.

The process of code graph generation and inference are the same as described in Section \ref{sec:trainimple}. Here, we will focus on the mapping from code graph to the original smart contract.
As illustrated below, mapping the node-level predictions back to the smart contract can be viewed as a reverse process of AST construction. We use the \texttt{predictLabel} function, which takes in the node predictions (\texttt{NodeLabel}), the source code at the byte level (\texttt{codeBit}), the source code length (\texttt{length}), and the AST's node attributes \texttt{src} list (\texttt{node2sourceCode}) as inputs to generate the line-level predictions (\texttt{LineLabel}).

\begin{lstlisting}[language=Python, numbers=none]
def predictLabel(node2sourceCode,NodeLabel,codeBit,length):
    labels=[0]*len(length)
    if sum(NodeLabel)==0:
        return [0]*len(length)

    for i,src in enumerate(node2sourceCode):
        src=src.split(":")
        sBit=codeBit[:int(src[0])]
        eBit=codeBit[:(sBit+int(src[1]))]

        start=len(sBit.decode("utf-8").splitlines())
        end=len(eBit.decode("utf-8").splitlines())
        codeLine=[*range(start, end+1,1)]
        
        if NodeLabel[i]>0:
            labels[codeLine] = 1
        else:
            labels[codeLine] = 0

    return labels
\end{lstlisting}

\section{Experiments}
\label{sec:experiment}
\subsection{Experiment Setup}
\sys is implemented with Python 3.8.10 and benchmarked on Linux Mint 20.3. Our workflow is built upon PyTorch~\cite{pytorch} version 1.9.0 and PyG~\cite{pytorchGeometric} version 2.0.4. The GNN model is trained on NVIDIA A16 graphic card consisting of four GPUs each with 16 GB RAM, and 4 CPUs with 128 GB RAM.

\subsection{Dataset}

To evaluate the performance of various vulnerability detection models, we employed the following three datasets:

\noindent\textbf{Peculiar Dataset}~\cite{Peculiar}: This dataset consists of 46,057 smart contracts with contract-level reentrancy vulnerability annotation for subtype \texttt{call}.

\noindent\textbf{\sys Dataset}: Our collected dataset of 13,773 contracts with line-level reentrancy vulnerability annotations for subtypes \texttt{call}, \texttt{transfer}, and \texttt{send}, as described in Section~\ref{sec:data}.

\noindent\textbf{Simulated Dataset}~\cite{Mando}: This dataset includes 31 smart contracts related to reentrancy vulnerability used by MANDO as their test set. The contracts consist of synthetic samples created by injecting vulnerabilities into clean smart contracts and public smart contracts. All samples include line-level annotations.

\subsection{Evaluation Metrics}
In the experiment, we benchmark the model performance at both line-level and contract-level. For line-level classification, we use \sys to determine if a node is vulnerable. For contract-level classification, after obtaining the classification result of each node, if there is a vulnerable node, we classify the contract as vulnerable; otherwise, it will be classified as non-vulnerable.

In order to evaluate the performance of different models, we compared their classification results with the annotated ground truth from Sec.\ref{sec:data}, and computed the confusion matrix at both the line-level and contract-level, including True Positive (TP), True Negative (TN), False Positive (FP), and False Negative (FN). We then used the following metrics to evaluate the performance of the models based on the confusion matrix:

\noindent\textbf{Accuracy (A)}: The proportion of correctly identified graphs or nodes, which is formulated as $\frac{TP + TN}{TP + FP + TN + FN}$.
    
\noindent\textbf{Recall (R)}: The proportion of correctly identified vulnerable graphs or nodes, which is formulated as $\frac{TP}{TP + FN}$.
    
\noindent\textbf{Precision (P)}: The proportion of identified vulnerable graphs or nodes that are actually vulnerable, which is formulated as $\frac{TP}{TP + FP}$.

\noindent\textbf{F1 score}: A harmonic mean that combines precision and recall, which is formulated as $\frac{2 * (P * R)}{(P + R)} $

\subsection{Baselines} 

For contract level classification, we compare \sysnospace's performance with the following baselines on Peculiar Dataset~\cite{Peculiar}. For line level classification, we compare \sys with MANDO on its simulated dataset.

\noindent\textbf{MANDO}~\cite{Mando}: It first uses a topology GNN to obtain node embeddings of a heterogeneous code graph and then classifies the vulnerability at the graph level. If the code graph is vulnerable, it trains a node classification model to identify the vulnerable nodes.

\noindent\textbf{DR-GCN}~\cite{GNNpaper1}: It first converts the contract into a symbolic graph and normalizes the graph by performing node elimination. The resulting code graph is classified using a degree-free GCN. 

\noindent\textbf{TMP}~\cite{GNNpaper1}: The graph processing step is the same as DR-GCN, but it uses a temporal message propagation network to classify code graphs.

\noindent\textbf{AME}~\cite{GNNpaper3}: It utilizes a rule-based approach to extract local expert patterns and convert the code into a semantic graph to extract global graph features. These two features are then fused using an attention-based network to predict vulnerabilities.

\noindent\textbf{Peculiar}~\cite{Peculiar}: It first extracts the dataflow of important source code variables and creates a crucial dataflow graph. The vulnerability is then classified using GraphCodeBert.

As mentioned earlier, SCScan~\cite{SCScan} also performs line-level vulnerability detection. However, at the time of submission, SCScan's code was not open-sourced, and their results were not reported on an open benchmark. Therefore, we could not compare their performance with that of \sys in this paper.

\subsection{Results on \sys Dataset} \label{sec:res_sys_dataset}
\begin{figure*}[ht]
  \begin{center}
   \subfloat[Subtype Call at node level]{\includegraphics[width=0.3\linewidth]{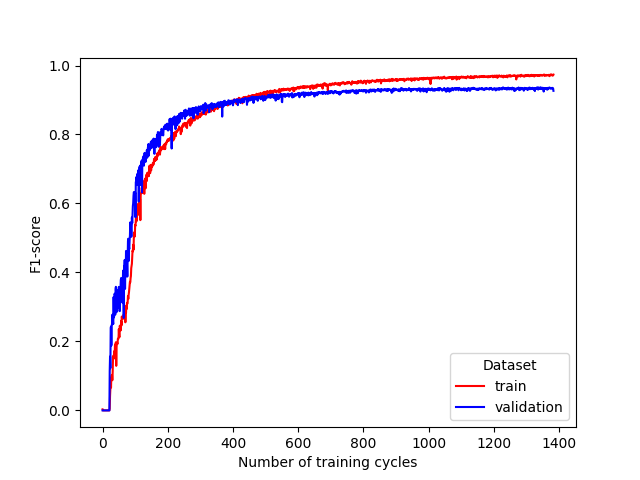}}
   \subfloat[Subtype Send at node level]{\includegraphics[width=0.3\linewidth]{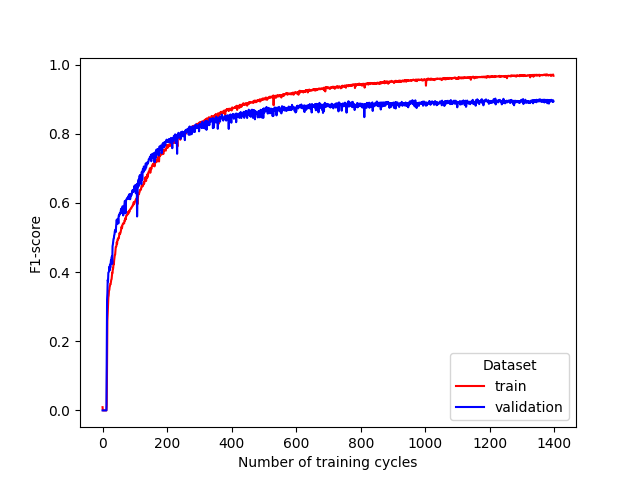}}
    \subfloat[Subtype Transfer at node level]{\includegraphics[width=0.3\linewidth]{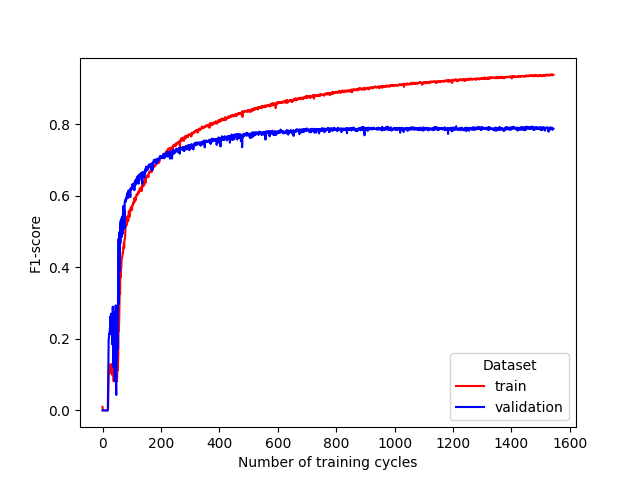}}  \\
    \subfloat[Subtype Call at graph level]{\includegraphics[width=0.3\linewidth]{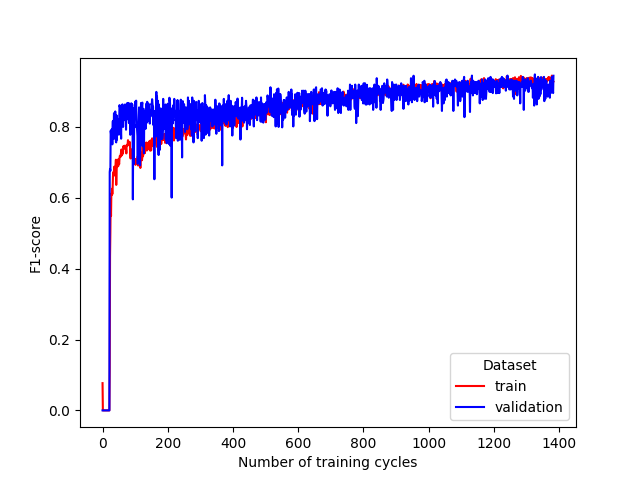}}
   \subfloat[Subtype Send at graph level]{\includegraphics[width=0.3\linewidth]{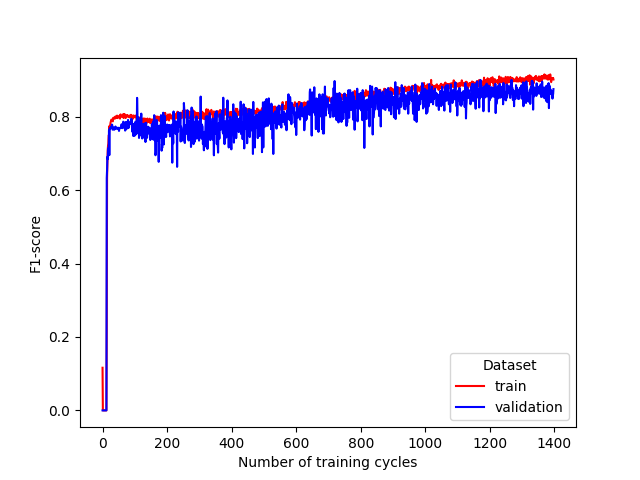}}
    \subfloat[Subtype Transfer at graph level]{\includegraphics[width=0.3\linewidth]{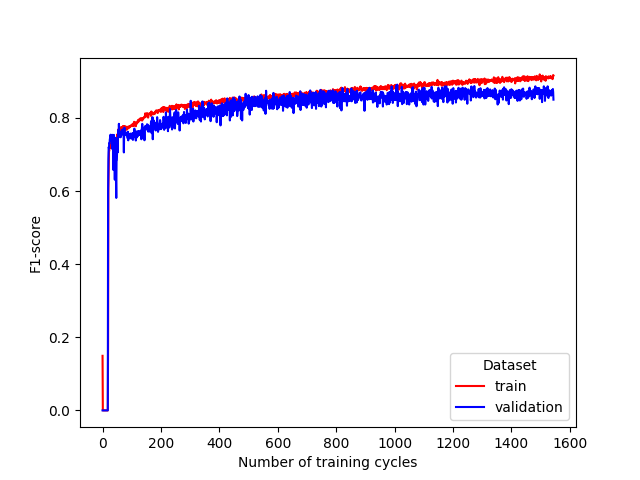}}
  \end{center}
  \caption{The F1 score evaluated during training for train and validation dataset on three subtypes at both node and graph level}
  \label{fig:f1_score}
  \vspace{-0.5cm}
\end{figure*}

We present the graph-level and node-level classification results on the \sys dataset's validation and test part in Table \ref{tab:result1}. The table displays the accuracy, precision, recall, and F1-score for three reentrancy subtypes: Call, Send, and Transfer. 

Based on the results presented in the table, we can make the following observations for node-level classification: Firstly, the classification accuracy for all datasets and reentrancy subtypes is above 99\%, indicating that \sys can successfully localize vulnerable nodes in real-world datasets. It demonstrated the effectiveness of our proposed pipeline. Secondly, node-level F1-score are higher than 90\% for most general vulnerable reentrancy subtype \texttt{Call}, and higher than 79\% for the rest two subtypes, showing \sys can benefit reentrancy vulnerability localization in real world smart contracts.

In terms of graph-level classification, we can find the following: Firstly, the classification accuracy for all reentrancy subtypes is higher than 89\%. This indicates that even though \sys is designed for line-level vulnerability localization, it can still effectively classify vulnerable contracts. Secondly, the F1-scores for all three reentrancy subtypes are higher than 85\%, demonstrating that \sys can successfully detect potential vulnerabilities.

\begin{table}[ht]
\centering
    \begin{tabular}{ccccccc}
    \toprule
 Subtype & Dataset & Level & Accuracy & Precision & Recall & F1-Score\\
       \midrule
 \multirow{4}{*}{Call} & validation set & node & 99.95\% & 94.88\% & 92.52\% & 93.69\% \\ 
		 &test set & node & 99.92\% & 93.23\% & 89.26\% & 91.20\% \\ 
		 &validation set & graph & 98.33\% & 87.59\% & 99.17\% & 93.02\% \\ 
		 &test set & graph & 97.31\% & 80.27\% & 100\% & 89.06\% \\ 
   \midrule
  \multirow{4}{*}{Send} & validation set & node & 99.86\% & 91.34\% & 89.05\% & 90.18\% \\ 
		&test set & node & 99.77\% & 92.55\% & 82.65\% & 87.32\% \\
	    &validation set & graph & 94.72\% & 81.07\% & 98.50\% & 88.94\% \\ 
		&test set & graph & 95.46\% & 80.07\% & 99.56\% & 88.76\% \\ 
   \midrule
  \multirow{4}{*}{Transfer}&validation set & node & 99.52\%  & 79.58\%  & 79.19\%  & 79.39\%  \\ 
		&test set & node & 99.53\%  & 83.61\%  & 78.92\%  & 81.19\%  \\ 
	    &validation set & graph & 89.91\%  & 74.71\%  & 99.69\%  & 85.41\%  \\ 
		&test set & graph & 90.74\%  & 78.21\%  & 98.55\%  & 87.21\%  \\ 
    \bottomrule
    \end{tabular}
    \caption{\sys performance on different reentrancy subtypes\label{tab:result1}}
\end{table}

In Figure~\ref{fig:f1_score}, we display the F1-scores on the train and validation datasets for the three reentrancy subtypes. From the figure, we observe that as the training progresses, \sys achieves an average F1-score of over 90\% on both the train and validation datasets for the most common subtype, \texttt{Call}, at both the node-level and graph-level. The average F1-scores for the subtypes \texttt{Transfer} and \texttt{Send} are also higher than 80\%. The stable curve indicates that \sys continuously learns from the train dataset and generalizes well toward the validation dataset.

\subsection{Comparison with Baselines}

\subsubsection{Contract-level Performance} 
In Table \ref{tab:comparison1}, we present the accuracy and F1-score of the contract-level vulnerability classification performance of \sysnospace, along with other baselines. However, since prior work reported their performance on the Peculiar Dataset, which only annotates reentrancy subtype \texttt{Call}, and some of them did not open-source their code, we were unable to benchmark them on our extended reentrancy vulnerability dataset in Section~\ref{sec:res_sys_dataset}. Therefore, we benchmark \sys on the Peculiar Dataset and report our results in the first row of the table. We note that prior works mainly constructed code graphs based on semantic information, whereas \sys relies on AST to generate code graphs. If a smart contract is missing its reference file, we cannot construct AST and perform inference on the contract, which is reasonable as one cannot determine the vulnerability without checking the entire contract code. Additionally, during \sys inference, we removed duplicated code graphs, which accounted for more than 50\% of the total. 

Based on the results, we observe that even when the duplicates are removed, \sys can still achieve 8\% better accuracy than the baseline methods. Although Peculiar~\cite{Peculiar} achieves a slightly better F1-score than \sysnospace, it cannot localize reentrancy vulnerabilities. Moreover, among the baseline methods, only MANDO performs fine-grained vulnerability localization, and \sys achieves a 21\% better F1-score at the contract level detection. 
It is noteworthy that even though \sys is designed to localize vulnerabilities at the node level, it still achieves good performance at the contract level detection.

\begin{table}[htbp]
  \centering
  \begin{tabular}{ccc}
    \toprule
    Method & Accuracy & F1-Score\\
    \midrule
    \sys & 98.15\% & 91.74\%\\ 
    \midrule
    MANDO \cite{Mando}  &  & 75.80\% \\
    DR-GCN \cite{GNNpaper1}  & 81.47\% & 76.39\% \\
    TMP \cite{GNNpaper1} & 84.48\% & 78.11\% \\
    AME \cite{GNNpaper3} & 90.19\% & 87.94\%\\
    Peculiar \cite{Peculiar}  & 99.81\% & 92.10\%\\
    \bottomrule  
  \end{tabular}
  \caption{Comparison with baselines on Peculiar Dataset.\label{tab:comparison1}}
\end{table}

\subsubsection{Node-level Performance}

 We compared \sys with MANDO on a simulated dataset since it was not possible to directly compare the two methods on our collected \sys Dataset. This was because MANDO did not provide explicit details on how their contract graph is mapped back to the smart contract, and line-level accuracy could not be compared. Additionally, directly comparing the two methods at the node level was not possible as MANDO used metapaths to construct the code graph while \sys used the AST to generate the code graph.

Therefore, to compare the two approaches, we use the simulated dataset collected by MANDO. 
One of the smart contracts could not be compiled, while another one belongs to the subtype \texttt{transfer}, which was correctly identified by \sysnospace.
As shown in Table~\ref{tab:comparison2}, \sys achieved 91.28\% F1-score on \texttt{call} reentrancy, outperforming its baseline MANDO. This result suggests that by incorporating code hierarchy information into graph modeling and training on real-world datasets, \sys outperformed MANDO in vulnerability localization. %
We also note that MANDO's effectiveness is heavily reliant on identifying vulnerabilities at the contract level, and its classification F1-score for certain vulnerability types, such as reentrancy, falls to 75.80\%.

\begin{table}[htbp]
  \centering
  \begin{tabular}{ccc}
    \toprule
    Method & Accuracy & F1-Score\\
    \midrule
    \sys & 97.44\% & 91.28\% \\ 
    Mando \cite{Mando}  & - & 86.40\% \\
    \bottomrule  
  \end{tabular}
  \caption{Comparison with MANDO on Simulated Dataset.\label{tab:comparison2}}
\end{table}

\subsection{Inference Overhead}

We present the inference overhead of smart contracts with varying lengths in Table~\ref{tab:inference}. The inference overhead is measured for five smart contracts with different Lines of Code (LoC) and includes the time it takes for AST construction, code graph generation, and GNN predictions. AST construction and code graph generation are performed on the CPU, while GNN prediction inference is performed on the GPU.

From the table, we can find the following: Firstly, the majority of the \sys overhead comes from AST construction using Solidity compilers. As the size of the smart contract increases, the percentage of time required for AST construction as a fraction of the total overhead increases. This is because the compiler requires more time to parse larger files. Secondly, despite an LoC value as large as 6.1k, the total inference time remains under 1.2s, demonstrating the efficiency of \sysnospace.

\begin{table}[htbp]
  \centering
  \begin{tabular}{ccccc}
    \toprule
    LoC & AST (ms) & CGG (ms) & Prediction (ms) & Total (ms)\\
    \midrule
    5 & 3.60 & 1.19 & 5.05 & 9.84 \\
    197 & 31.29 & 6.04 & 9.96 & 47.29\\ 
    245 & 55.89 & 12.92 & 14.15 & 82.96\\ 
    5815 & 901.98 & 178.08 & 95.26 & 1175.32\\ 
    6151 & 850.70 & 191.48 &  108.25 & 1150.43\\
    \bottomrule  
  \end{tabular}
  \caption{\sys overhead on AST construction (AST), code graph generation (CGG) and prediction\label{tab:inference}}
\end{table}
\vspace{-0.5cm}
\section{Related Work}
\subsection{Smary Contract Vulnerability Detection}

\textbf{Symbolic Execution:} Symbolic execution~\cite{Osiris,SmarTest,SmartCheck,teEther,SESCon} takes the smart contracts bytecode as input and uses symbolic expression to represent the contracts. Then, it uses Satisfiability Modulo Theory (SMT) solver to prove if certain vulnerability conditions can exist in the source code or not. In smart contract vulnerability detections, symbolic executions are always combined with other rule-based algorithms to detect potential security risks. 
Osiris \cite{Osiris} detects integer bugs with symbolic execution and additional taint analysis, which is a kind of tracking of the data across the control flow. 
SmarTest \cite{SmarTest} applies symbolic execution on the smart contract language models and creates transaction sequences to detect bugs.
SAILFISH \cite{SAILFISH} uses a hybrid approach for source codes, where a storage dependence graph performs analysis of side effects on the storage variables during the execution and follows by the symbolic evaluation to detect vulnerabilities. 

\textbf{Fuzzing:} Fuzzing~\cite{ContractFuzzer,Echidna,FuzzingEstimateGasCosts,ConFuzzius,MTFuzz,Neuzz} uses different algorithms to automatically generate inputs  potentially trigger errors or unexpected behaviors to detect smart contract  vulnerabilities. 
ContractFuzzer~\cite{ContractFuzzer} analyzes the ABI interfaces of smart contracts to generate inputs that conform to the invocation grammars of the smart contracts under test. It defines new test oracles for different types of vulnerabilities and instrument EVM to monitor smart contract executions to detect security vulnerabilities. 
Echidna~\cite{Echidna} incorporates a worst-case gas estimator into a general-purpose fuzzer. When a property violation is detected, a counterexample is automatically minimized to report the sequence of transactions that triggers the failure.

\textbf{Machine Learning:} Machine training~\cite{SC-VDM,ContractWard,Giesen2022PracticalMO} based algorithms take the smart contracts as text data or graph data and use different machine learning algorithms to classify if a smart contract is vulnerable or not. The machine learning algorithms enable the model to learn the hidden feature representation of smart contract code that previously introduced rule-based algorithms failed to catch. Most of them~\cite{SC-VDM,GNNpaper1,Peculiar} classify vulnerabilities at contract level, and others first perform contract level vulnerabilities detection and then localize the bugs at line level like MANDO~\cite{Mando} and SCScan~\cite{SCScan}. SC-VDM \cite{SC-VDM} converts the bytecode into a greyscale image and applies a convolutional neural network to classify whether a smart contract is vulnerable. Liu et al.~\cite{GNNpaper1} propose to apply a graph convolutional neural network on a contract graph of the source code to detect bugs by graph classification. In the follow-up papers, they \cite{GNNpaper2,GNNpaper3} 
integrate additional expert patterns to increase detection accuracy. Peculiar \cite{Peculiar} extracts the dataflow of important variables of the source code in a crucial dataflow graph. Then, the source code and the dataflow graph are fed into GraphCodeBERT \cite{GraphCodeBert} for reentrancy vulnerability classification.

\subsection{Source Code Graphs}
In machine learning based vulnerability detection algorithms, converting source code into image or texts may lose structural information. Therefore, many work first convert the source code into graph structure, and then classify the graph using different GNN models.

\textbf{Heterogeneous Code Graph} Heterogeneous code graph~\cite{GNNpaper1,GNNpaper2,GNNpaper3,Mando} construct graph connections based on the semantic information and program flow paths, which is widely used in the smart contract vulnerability detection.  However, they failed to incorporate code dependencies and hierarchies into modeling. 
Liu et al.~\cite{GNNpaper1,GNNpaper2,GNNpaper3} proposed to construct code graph based on three different node types. Major nodes represent important function invocations or critical variables. Other variables are seen as secondary nodes and fallback nodes symbolize a fallback function call. Edges are defined by the feasible program flow paths. Additionally, an elimination step is performed to reduce the graph size, which removes secondary and fallback nodes and redirects the edges. This final normalized graph is fed into a GNN that detects reentrancy, timestamp dependence, and infinite loop vulnerabilities.
MANDO~\cite{Mando}'s code graphs are based on call graphs and control flows of the source code. It first uses a multi-metapaths extractor to generate metapaths from the node types and their associated edges in these graphs. The node embeddings are created with these metapaths and fused with the metapaths by a heterogeneous attention mechanism at node level. Afterward, these node embeddings are fed into an MLP for graph classification to check whether the contract is vulnerable. If that is the case, the exact location of the vulnerability inside the contract is determined by node classification with the updated node embeddings in graph classification.

\textbf{AST based Code Graph} AST~\cite{Allamanis,Allamanis2021,CPG} based code graph generation is widely used in code vulnerability detection like C/C++. AST can extract better structure and variable/function dependency from source code. Some recent work ~\cite{xu2021novel,yang2021multi} also applied AST based graph analysis to smart contract vulnerability detection. 
Allamanis et al.~\cite{Allamanis,Allamanis2021} proposed a program graph based on the node connections in \Csharp's AST. It first transforms the \Csharpspace source code into AST representation using compilers. Then, the nodes in the tree are replaced by corresponding source code tokens and connected with edges that reflect the original execution order. To represent the control and dataflow structure, they add additional edges, for example, between all occurrences of the same variable or in condition statements to indicate all valid paths. 
Code Property Graphs~\cite{CPG} proposed to construct graphs from the C/C++'s AST by combining it with edges from control flow and program dependence. The control flow edges are inserted for subsequent statements, loops, returns, and similar constructs.  The program dependence edges are inserted to reflect the influences of other variables or predicates on a certain variable, such as the definition or variable assignment.

\section{Conclusion and Future Work}
\label{sec:conclusion}

In the paper, we present \sys, the first end-to-end fine-grained line-level reentrancy vulnerability detection system with evaluations on the first-of-its-kind real world dataset. \sys first converts source code smart contracts to code graphs, and then trains node classification models on the graphs to localize vulnerabilities. Our experiments demonstrate that \sys achieves 93.02\% F1-score in contract-level vulnerability detection and 93.69\% F1-score in line-level vulnerability localization on the real world dataset. Moreover, \sys processes low inference overhead and can localize vulnerabilities within less than 1.2s for 6k LoC smart contracts. We will open-source \sys along with its dataset to promote research in this area.

In the future, we plan to extend our real world dataset to include multiple vulnerability annotations and explore different code graph representations that may improve GNN classification performance.


\bibliographystyle{IEEEtranS}
\bibliography{refs,bibliography}


\end{document}